# Studies on the Origin and Evolution of Codon Bias

*Jan C Biro*[1]

Homulus Foundation, 612 S Flower St., #1220, Los Angeles, 90017 CA, USA
Jan.biro@att.net
www.janbiro.com

[1]Corresponding author

## Abstract

**Background:** There is a 3-fold redundancy in the Genetic Code; most amino acids are encoded by more than one codon. These synonymous codons are not used equally; there is a Codon Usage Bias (CUB). This article will provide novel information about the origin and evolution of this bias.

**Results:** Codon Usage Bias (CUB, defined here as deviation from equal usage of synonymous codons) was studied in 113 species. The average CUB was 29.3 ± 1.1% (S.E.M, n=113) of the theoretical maximum and declined progressively with evolution and increasing genome complexity. A Pan-Genomic Codon Usage Frequency (CUF) Table was constructed to describe genome-wide relationships among codons. Significant correlations were found between the number of synonymous codons and (i) the frequency of the respective amino acids (ii) the size of CUB. Numerous, statistically highly significant, internal correlations were found among codons and the nucleic acids they comprise. These strong correlations made it possible to predict missing synonymous codons (wobble bases) reliably from the remaining codons or codon residues.

**Conclusions:** The results put the concept of "codon bias" into a novel perspective. The internal connectivity of codons indicates that all synonymous codons might be integrated parts of the Genetic Code with equal importance in maintaining its functional integrity.

# Background

The genetic code is redundant: 20 amino acids plus start and stop signals are coded by 64 codons. This redundancy increases the resistance of genes to mutation: the third codon letters (wobble bases) can often be interchanged without affecting the primary sequence of the protein product. Nevertheless, wobble base usage is highly conserved in mRNA sequences (there is no or very little individual or intra-species variation) and, interestingly, some wobble mutations (though they are called silent mutations) are known to cause genetic disease with no change in the amino acid sequences [1].

However, the wobble bases are not randomly selected, as they might be if interchangeability were unrestricted. There is codon bias, i.e. codon usage is not equally distributed between the possible synonyms; some redundant codons are preferentially used. This bias is described in Codon Usage Frequency (CUF) Tables [2].

Many studies confirm the existence of codon bias and significant correlations have been found between codon bias and various biological parameters such as gene expression level [3-6] gene length [7-9], gene translation initiation signal [10], protein amino acid composition [11], protein structure [12-13], tRNA abundance [14-17], mutation frequency and pattern, [18-19] and GC composition [20-23].

These observations may not be universally valid because some statistically significant observations in one species are not reproduced in another. However, there is a strong expectation that codon bias, which is obviously well conserved in different species, reflects a general biological function because of the universal nature of the Genetic Code and the structure and function of nucleic acids and proteins.

The aim of this study is to investigate the possible origin of so-called "codon bias", measure it quantitatively and compare it among many species.

# Materials and methods

Codon Usage Frequency (CUF) Tables were obtained for 113 different organisms from the Codon Usage Database (NCBI-GenBank, update: November 16, 2006 [24]). The organisms were selected from KEGG (Kyoto Encyclopedia of Genes and Genomes, [25]) and represented a wide variety of species from different evolutionary lines [**Supplementary File 1**].

To calculate Codon Usage Bias (CUB) numerically, I assumed that statistically equal usage of all available synonymous codons is the neutral "starting point" for the development of species-specific codon usages, and the CUB is the sum of the deviations from such random, equal usage.

The codons (i, 64) were divided into 21 subgroups (j, corresponding to the 20 amino acids and 1 stop signal). The number of occurrences of a codon was normalized and the frequencies of the codons ($CUF_{ij}$) in each fraction were calculated. The sum of $CUF_{if}$ in a fraction was always treated as 100% so the sum of all fractions was 2100%. $n_i$ is the number of synonymous codons in the $j^{th}$ fraction and $n_j=64$

$$\sum_{i=1}^{n_j} CUF_{ij} = 100 \ (\%) \qquad \sum_{j=1}^{n_j} CUF_{ij} = 2100 \ (\%)$$

$CUF_{ij}$ is the frequency (%) of the $i^{th}$ codon in the $j^{th}$ fraction encoded by $n_i$ synonymous codons.

These fractional frequencies were compared to the random fractional frequencies ($rCUF_{ij}$), defined as the fractional frequency that a codon would have if all alternative codons were used randomly and equally.

$$rCUF_{(1j)} = rCUF_{(2j)} = rCUF_{(n)j} = rCUF_{(ij)} = 100/n_i \quad (\%)$$

The sum of rCUF in a fraction is also 100% and in each fraction altogether is 2100%.

CUB is defined as the absolute difference between CUF and rCUF:-

$$CUB_{ij} = |\ CUF_{ij} - rCUF_{ij}\ | \quad \text{and}$$

$$CUB = \sum_{i=1}^{64} |\ CUF_{ij} - rCUF_{ij}\ | \quad (\%)$$

More simply, CUB is the absolute number of fractional frequencies minus the number expected if usage of synonymous codons was uniform.

CUB may be used in some cases with its +/- orientation indicated. In these cases, positive values indicate over-utilization of a codon (e.g. dominant codons) while negative values indicate under-utilization (suppression).

$CUB_{min} = 0$ if $CUF_{ij} = rCUF_{ij}$ and the Calculated Maximal Possible $CUB_{max}$ is 2416.7%. This is the value when only one of all the possible synonymous codons is used (100% frequency) for every amino acid and for the stop signal.

Further explanation of the CUB calculation is given in [**Supplementary File 2**], together with an example. $CUF_{ij}$ (%) is not to be confused with a "regular" codon frequency ($CUF_i$), which indicates the frequency of a codon in the entire genome (all 21 fractions) and is usually given in the CUF Tables in #/1000 units.

The definition of CUB in this article is not directly comparable to other widely used definitions such as CUI.

## Results

### Quantitative evaluation of codon bias

CUB = 0% when all available synonymous codons are equally used. The maximal calculated bias, $CUB_{max}$ = 100 %, indicates that only one codon is used for each amino acid (and for the stop signal), while the remaining 43 codons are not used at all. I calculated CUB in 113 species and found that the average value is 29.3+/-1.1% (S.E.M, n=113). There seems to be a modest but significant decrease in the bias during evolution: bacteria and archeoata have the highest bias while vertebrates have the lowest. Eukaryotes have significantly lower CUB than prokaryotes. Humans have the lowest value (18.9%) (Figure 1).

**Figure 1**

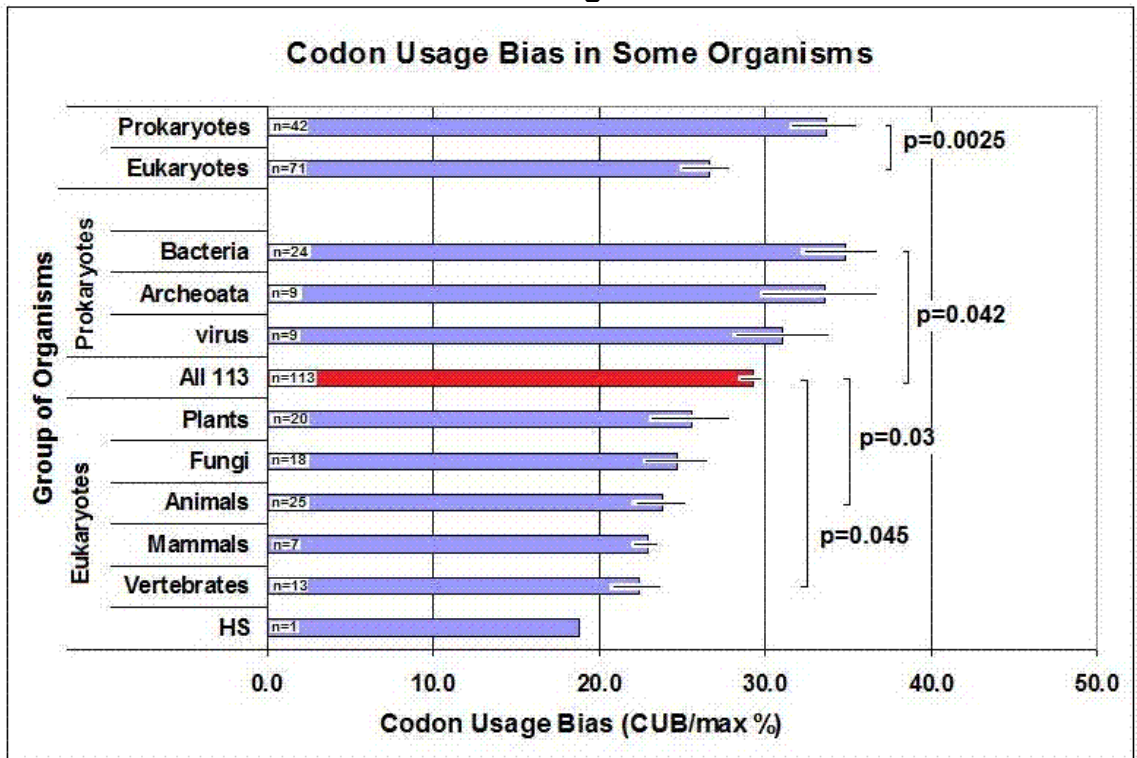

**Figure 1.: Codon Usage Bias (CUB) in Some Organisms.**
Mean+/- S.E.M, n: number of species in the group.

There is a slight negative correlation between the size of the codon- and gene-pool of an organism and its CUB (p<0.01, n=113, not shown). The size and complexity of both genome and proteome increase with evolution, while the CUB decreases. A larger codon pool seems to utilize more codon variation, which leads to lower differences between the usage frequencies of synonymous codons.

## Qualitative evaluation of CUB

Detailed analysis of different species reveals wide variations in CUB (Figure 2). There is a seemingly random variation in CUB between amino acids and different groups of organisms. However, a comparison of closely-related species with large codon pools shows very similar patterns. For example, all mammals have very similar CUB patterns.

**Figure 2**

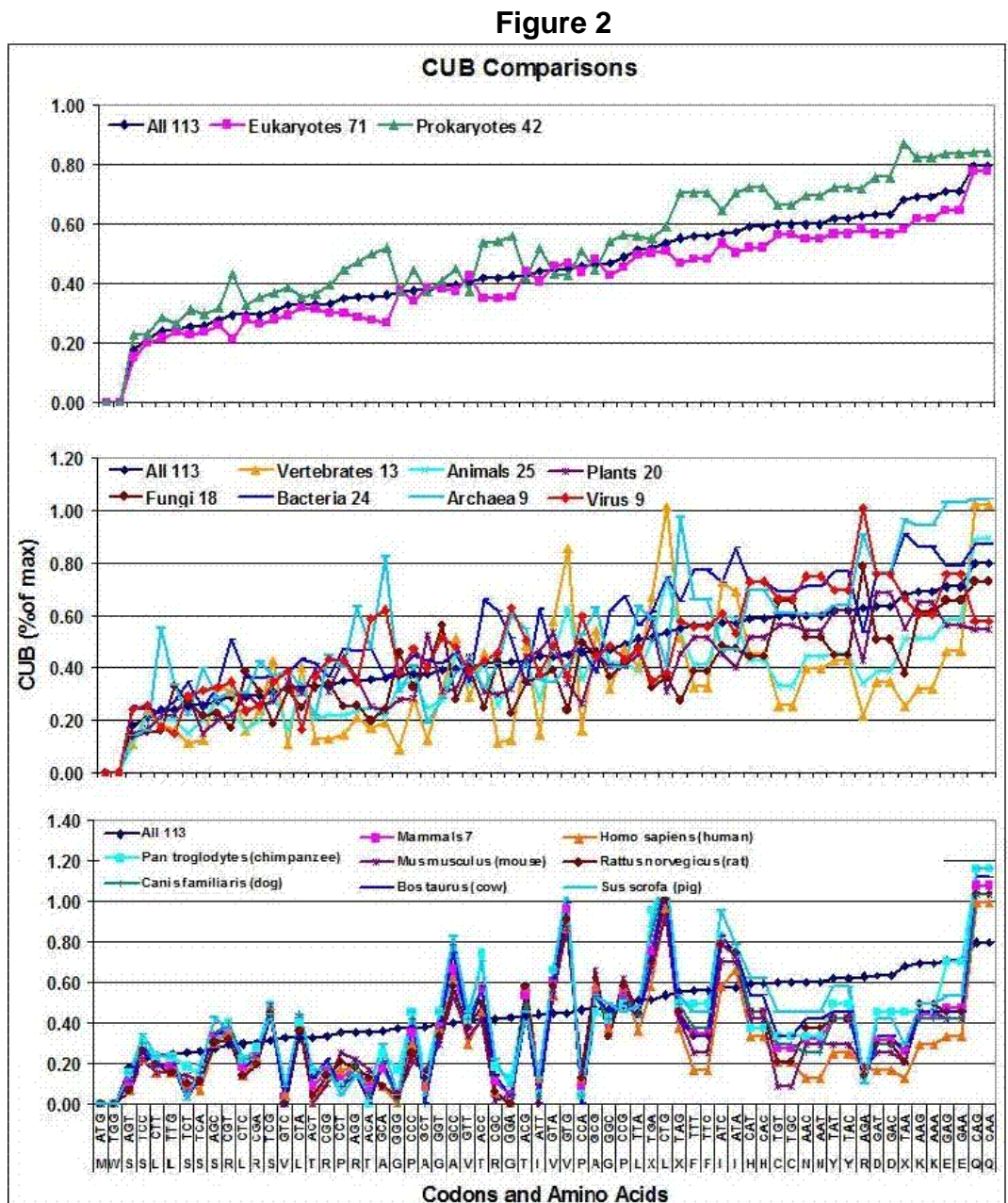

**Figure 2 CUB Comparisons.**
Codon Usage Biases (CUB) were calculated in 113 species and sorted into subgroups. The mean CUBs of the 64 codons in the indicated subgroups are shown. ($CUB_{max}$=100% for the 64 cadons altogether).

## Pan-genomic codon usage

I accumulated the CUF data from the 113 species into a single CUF Table (Table I).

### Table I

| Am.Acid | Codon | Number | CUFi (#/1k) | CUFij (% of fraction) | rCUF (% of fraction) | CUBij (%) | |CUBij (%)| |
|---|---|---|---|---|---|---|---|
| Gly | GGG | 3598776.0 | 12.5 | 19.0 | 25.0 | -6.0 | 6.0 |
| Gly | GGA | 5477754.0 | 19.0 | 28.9 | 25.0 | 3.9 | 3.9 |
| Gly | GGT | 4451391.0 | 15.4 | 23.5 | 25.0 | -1.5 | 1.5 |
| Gly | GGC | 5445255.0 | 18.9 | 28.7 | 25.0 | 3.7 | 3.7 |
| Glu | GAG | 9756293.0 | 33.8 | 51.4 | 50.0 | 1.4 | 1.4 |
| Glu | GAA | 9209632.0 | 31.9 | 48.6 | 50.0 | -1.4 | 1.4 |
| Asp | GAT | 8195141.0 | 28.4 | 54.9 | 50.0 | 4.9 | 4.9 |
| Asp | GAC | 6731842.0 | 23.3 | 45.1 | 50.0 | -4.9 | 4.9 |
| Val | GTG | 6428801.0 | 22.3 | 35.3 | 25.0 | 10.3 | 10.3 |
| Val | GTA | 2695055.0 | 9.3 | 14.8 | 25.0 | -10.2 | 10.2 |
| Val | GTT | 4781792.0 | 16.6 | 26.3 | 25.0 | 1.3 | 1.3 |
| Val | GTC | 4280999.0 | 14.8 | 23.5 | 25.0 | -1.5 | 1.5 |
| Ala | GCG | 3487704.0 | 12.1 | 16.8 | 25.0 | -8.2 | 8.2 |
| Ala | GCA | 5031084.0 | 17.4 | 24.3 | 25.0 | -0.7 | 0.7 |
| Ala | GCT | 5779334.0 | 20.0 | 27.9 | 25.0 | 2.9 | 2.9 |
| Ala | GCC | 6432441.0 | 22.3 | 31.0 | 25.0 | 6.0 | 6.0 |
| Arg | AGG | 3071603.0 | 10.6 | 18.9 | 16.7 | 2.3 | 2.3 |
| Arg | AGA | 3953550.0 | 13.7 | 24.4 | 16.7 | 7.7 | 7.7 |
| Ser | AGT | 3473736.0 | 12.0 | 15.1 | 16.7 | -1.6 | 1.6 |
| Ser | AGC | 4391636.0 | 15.2 | 19.1 | 16.7 | 2.4 | 2.4 |
| Lys | AAG | 8869890.0 | 30.7 | 52.7 | 50.0 | 2.7 | 2.7 |
| Lys | AAA | 7946577.0 | 27.5 | 47.3 | 50.0 | -2.7 | 2.7 |
| Asn | AAT | 6514892.0 | 22.6 | 51.9 | 50.0 | 1.9 | 1.9 |
| Asn | AAC | 6036774.0 | 20.9 | 48.1 | 50.0 | -1.9 | 1.9 |
| Met | ATG | 6909100.0 | 23.9 | 100.0 | 100.0 | 0.0 | 0.0 |
| Ile | ATA | 3373624.0 | 11.7 | 22.2 | 33.0 | -10.8 | 10.8 |
| Ile | ATT | 5925942.0 | 20.5 | 39.0 | 33.0 | 6.0 | 6.0 |
| Ile | ATC | 5905801.0 | 20.5 | 38.8 | 33.0 | 5.8 | 5.8 |
| Thr | ACG | 2486009.0 | 8.6 | 15.9 | 25.0 | -9.1 | 9.1 |
| Thr | ACA | 4473401.0 | 15.5 | 28.6 | 25.0 | 3.6 | 3.6 |
| Thr | ACT | 4084032.0 | 14.2 | 26.1 | 25.0 | 1.1 | 1.1 |
| Thr | ACC | 4619047.0 | 16.0 | 29.5 | 25.0 | 4.5 | 4.5 |
| Trp | TGG | 3675912.0 | 12.7 | 100.0 | 100.0 | 0.0 | 0.0 |
| End | TGA | 308407.0 | 1.1 | 39.9 | 33.0 | 6.9 | 6.9 |
| Cys | TGT | 2509240.0 | 8.7 | 47.2 | 50.0 | -2.8 | 2.8 |
| Cys | TGC | 2810369.0 | 9.7 | 52.8 | 50.0 | 2.8 | 2.8 |
| End | TAG | 183171.0 | 0.6 | 23.7 | 33.0 | -9.3 | 9.3 |
| End | TAA | 281718.0 | 1.0 | 36.4 | 33.0 | 3.4 | 3.4 |
| Tyr | TAT | 4107194.0 | 14.2 | 48.6 | 50.0 | -1.4 | 1.4 |
| Tyr | TAC | 4337253.0 | 15.0 | 51.4 | 50.0 | 1.4 | 1.4 |
| Leu | TTG | 4737403.0 | 16.4 | 17.5 | 16.7 | 0.8 | 0.8 |
| Leu | TTA | 3136971.0 | 10.9 | 11.6 | 16.7 | -5.1 | 5.1 |
| Phe | TTT | 5426287.0 | 18.8 | 48.0 | 50.0 | -2.0 | 2.0 |
| Phe | TTC | 5873939.0 | 20.4 | 52.0 | 50.0 | 2.0 | 2.0 |
| Ser | TCG | 2485290.0 | 8.6 | 10.8 | 16.7 | -5.9 | 5.9 |
| Ser | TCA | 3962926.0 | 13.7 | 17.2 | 16.7 | 0.6 | 0.6 |
| Ser | TCT | 4499282.0 | 15.6 | 19.6 | 16.7 | 2.9 | 2.9 |
| Ser | TCC | 4191190.0 | 14.5 | 18.2 | 16.7 | 1.6 | 1.6 |
| Arg | CGG | 2379693.0 | 8.2 | 14.7 | 16.7 | -2.0 | 2.0 |
| Arg | CGA | 1898114.0 | 6.6 | 11.7 | 16.7 | -5.0 | 5.0 |
| Arg | CGT | 2071451.0 | 7.2 | 12.8 | 16.7 | -3.9 | 3.9 |
| Arg | CGC | 2842349.0 | 9.8 | 17.5 | 16.7 | 0.9 | 0.9 |
| Gln | CAG | 6974185.0 | 24.2 | 58.6 | 50.0 | 8.6 | 8.6 |
| Gln | CAA | 4922495.0 | 17.1 | 41.4 | 50.0 | -8.6 | 8.6 |
| His | CAT | 3408853.0 | 11.8 | 49.7 | 50.0 | -0.3 | 0.3 |
| His | CAC | 3453004.0 | 12.0 | 50.3 | 50.0 | 0.3 | 0.3 |
| Leu | CTG | 7327412.0 | 25.4 | 27.1 | 16.7 | 10.4 | 10.4 |
| Leu | CTA | 2418342.0 | 8.4 | 8.9 | 16.7 | -7.7 | 7.7 |
| Leu | CTT | 4540618.0 | 15.7 | 16.8 | 16.7 | 0.1 | 0.1 |
| Leu | CTC | 4907197.0 | 17.0 | 18.1 | 16.7 | 1.5 | 1.5 |
| Pro | CCG | 2861706.0 | 9.9 | 18.9 | 25.0 | -6.1 | 6.1 |
| Pro | CCA | 4491106.0 | 15.6 | 29.7 | 25.0 | 4.7 | 4.7 |
| Pro | CCT | 4142534.0 | 14.4 | 27.4 | 25.0 | 2.4 | 2.4 |
| Pro | CCC | 3615638.0 | 12.5 | 23.9 | 25.0 | -1.1 | 1.1 |
| | | | | 0.0 | 0.0 | 0.0 | 0.0 |
| Summs | | 288600157.0 | 1000.0 | 2100.0 | 2097.9 | 2.1 | 245.4 |

10.14% of CUBmax

This Table is intended to give a virtual representation of all organisms (Pan-Genome) and a numerical representation of the "universal" translation machinery. As many as 288xE10 codons are represented in this collection.

The distribution of CUB values in the Pan-Genomic CUF Table is illustrated in Figure 3. The transition from maximum-positive to maximum-negative values is smooth and there is no obvious or unambiguous border between the so-called *dominant* and *prohibited* codons. All possible codons are used.

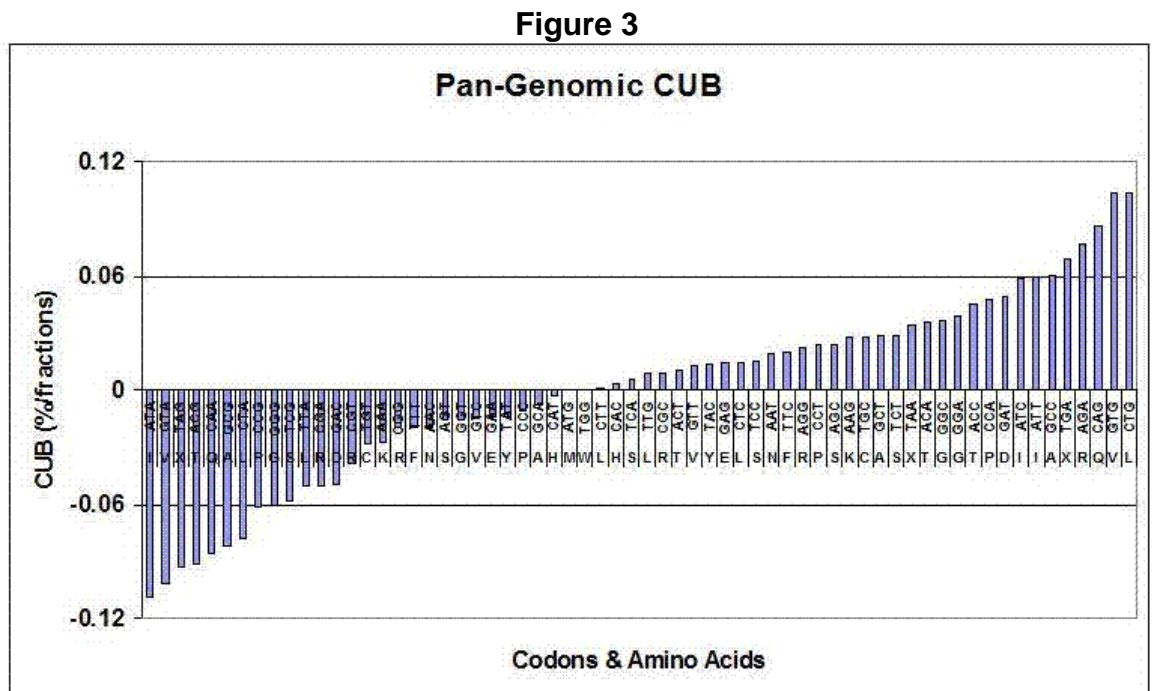

**Figure 3.: Distribution of Pan-Genomic CUB**
CUB was taken from Pan-Genomic Codon Usage Table and sorted in ascending order.

There is a significant positive correlation between the number of synonymous codons ($n_i$, #/amino acid) and the propensity of amino acids in the proteome (#/1000 amino acid residues). A similar correlation exists between synonymous codon frequency and CUB (Figure 4). These important correlations were discovered by analyzing the Pan-Genomic CUF Table (64 values) and were confirmed using individual data from all species (113x21 values).

Another possible way to evaluate the possible phylogenetic relationships among CUBs in different species is to use the Pan-Genomic CUB Table as a common reference. I performed correlation analyses and compared the lists of species-specific CUB values to the list of mean CUB values in the Pan-Genomic CUB Table (64x113 comparisons), then used the significance of correlations as an indicator of CUB distances [**Supplementary File 3**].

**Figure 4**

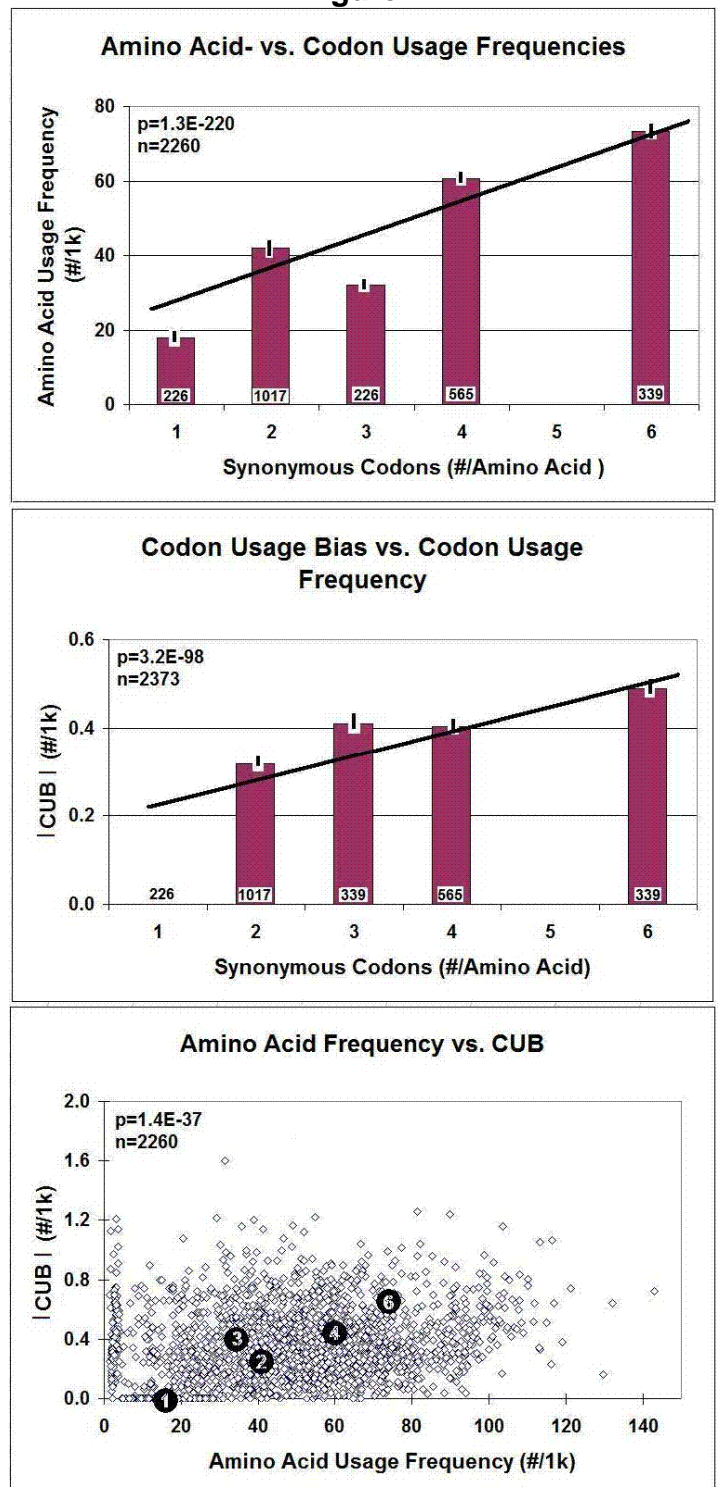

**Figure 4.: Correlations between Synonymous Codon Usage Frequency, Amino Acid Usage Frequency and Codon Usage Bias (CUB).**

The columns represent mean ± S.E.M., n is indicated within the columns. The significance of correlations is also included. Black circles indicate the positions of mean values and the numbers in the black circles indicate the number of synonymous codons/amino acid.

I found that the CUB of vertebrates is most similar (least distant) to the average CUB, while bacteria and viruses are most distant from it. This correlation analysis involves all codons and gives no information about the development of individual CUBs. I therefore compared the codon-specific CUB values in the 113 species to obtain a rough estimate of the stability of (commitment to) a CUB through evolution. The mean/SD of the 113 amino acid-specific CUB values gives a good estimate how this stability (Figure 5).

**Figure 5**

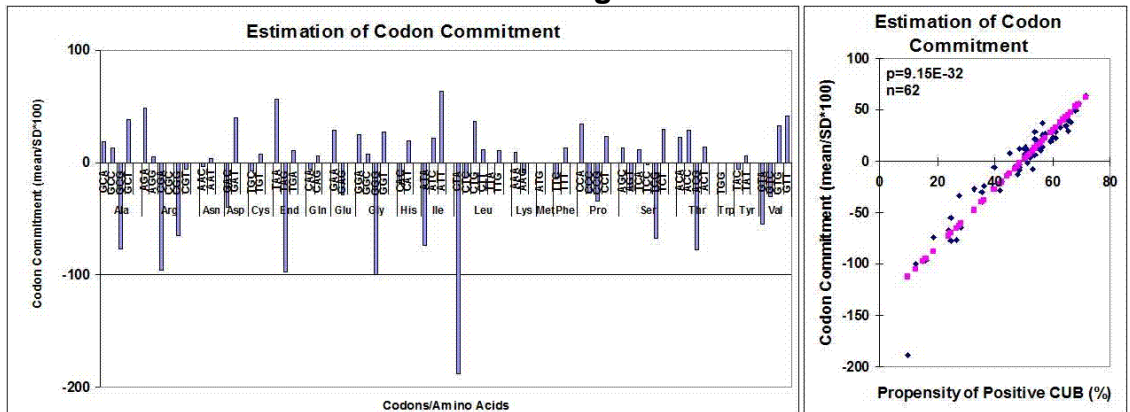

**Figure 5.: Estimation of Codon Commitment**
The mean ± SD values were calculated for the 64 codons (n=113). The mean/SD*100 values were regarded as the measure of a codon's commitment to a given CUB through evolution. Very low (-) values indicate strong negative CUB (under-utilization of that codon) while the meaning of high (+) values is the opposite. The codon commitment value reflects the propensity towards over-utilized codons (positive CUB).

**Internal dynamics of codons**

*Correlations between individual CUB frequencies*

When one of the synonymous codons is used more frequently than expected (positive CUB), another will be less frequently used (negative CUB). More generally, this means that codon usage changes in a subgroup of the 64 codons will be accompanied by changes in the opposite direction in the remaining codons.

I sorted the CUB values (64x113=7,232 listed in total) in the **Pan-Genomic CUB Table** according to their sizes and +/- directions [**Supplementary File 4**]. This sorting divided the 64 codons (c) into two subgroups (Ac and Bc) and the 113 species (s) into two additional groups (As and Bs). The Ac-As and Bc-Bs subgroups contained predominantly over-represented (positive CUB) codons and are located in the opposite diagonal corners of the Table. The Ac-Bs and Bc-As fields contained predominantly under-represented (negative CUB) codons and are located in the other opposite diagonal corners of the Table.

There is an internal inverse relationship between codons, which is valid and the same for all species. This inverse relationship is shown in a compressed and simplified form in Figure 6a, b.

**Figure 6**

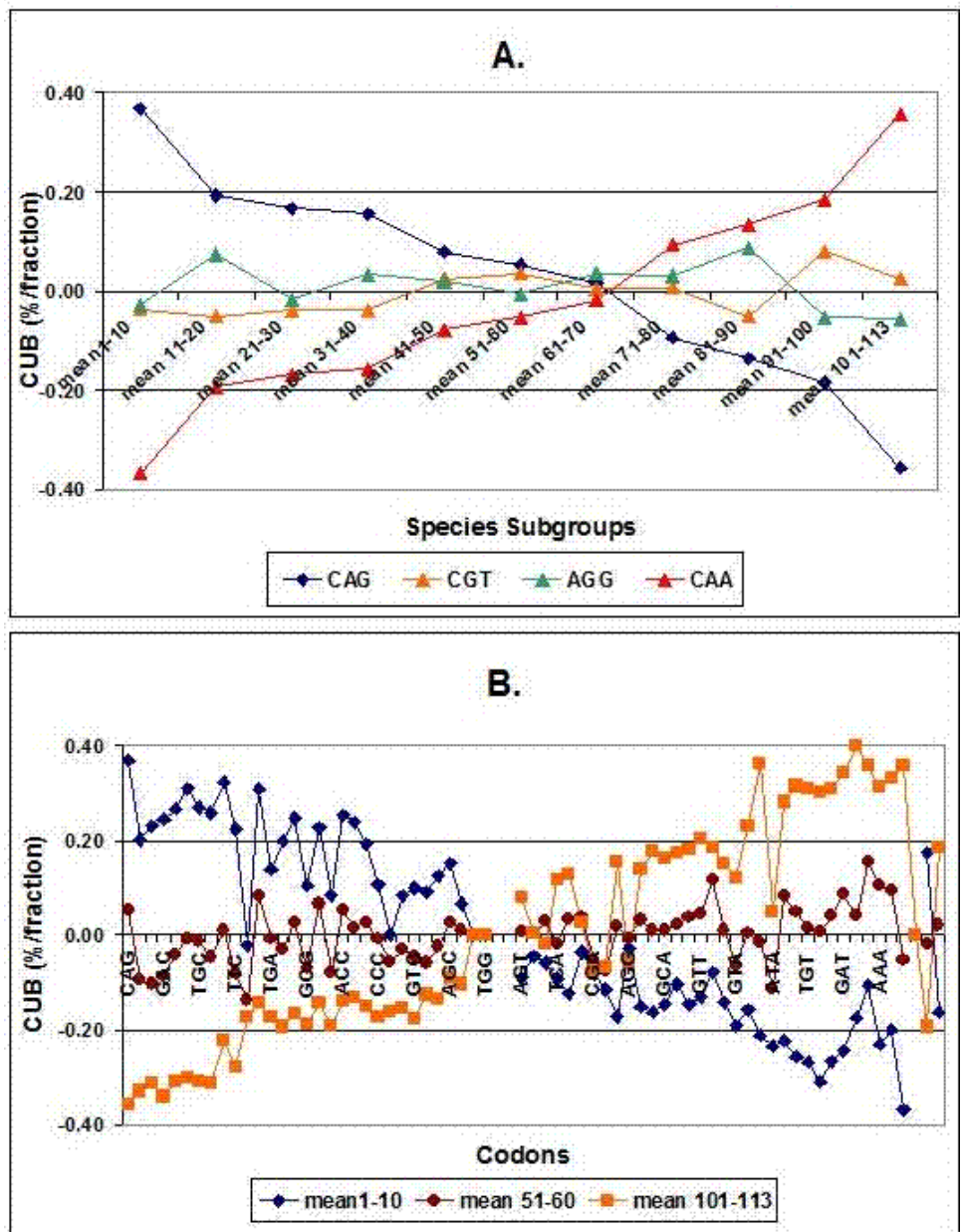

**Figure 6.: Species Dependent Internal Correlation between CUBs.**
Codon usage biases (CUBs) from 113 species were sorted as described in the text and divided into 11 consecutive subgroups. Each symbol represents the mean of CUB values from 10 different species. The values were sorted for species subgroups (A) and for codons (B). Only some representative samples are included (4 codons of total 64 and 3 groups of different species of total 11).

Negative correlations were expected between some subgroups of CUBs and others in the same species. Surprisingly, however, all codons and all species belong to only 2 clusters with highly correlated, opposite dynamics.

The above figures indicate that there is a close internal and inverse correlation between the CUBs of different codons. The magnitude and orientation of a CUB shows wide variation between species. Our collection of 113 species is too limited for any conclusion about the phylogenetic rules of development of CUB to be drawn, but the first impression is an absence of phylogenetic rules:
- about half the species under-utilize about half the codons, while the other half show the opposite behavior in respect of the remaining codons.
- It is difficult to find a correlation between CUB and taxon boundaries. All mammals (in the table) show a homogenous CUB pattern, while other taxa are much more diverse.
- Most codons show a wide pangenomic variation in CUB, but some vary much less than others (Figure 5). Some codons (TAG, GGG, CGA, CTA) are under-utilized by more than 80% of the 113 species listed, i.e. these synonymous codons have become committed to a given CUB orientation while others have not. There is a significant negative correlation between the proportion of codons committed to a given CUB orientation and the extent to which CUB varies (also apparent in Figure 5).

*Internal relationship among codon bases in codon usage tables*

Codons are defined by 3 nucleotides. Therefore, CUF Tables can be further analyzed as Nucleotide Usage Frequency (NUF) Tables.

The 113 CUF Tables in our material are based on 288 million codons and 690 K CDS. The number of codons in this collection is enough to provide reliable information about the general rules, if any, that determine nucleotide ratios and correlations in genomes.

There are some highly significant correlations among codon bases.
The fractional frequency of each nucleotide base in every codon position correlates positively with its complementary codon (Table 2).

**Table 2**
**POSITIONAL NUCLEOTIDE USAGE FREQUENCIES IN 113 SPECIES**

| log (-C) | C1/# | C2/# | G3/# | C3/# | G1/# | G2/# | T2/# | T1/# | T3/# | A3/# | A2/# | A1/# |
|---|---|---|---|---|---|---|---|---|---|---|---|---|
| A1/# | -45.1 | -31.7 | -29.2 | -26.5 | -24.3 | -20.6 | 5.5 | 16.6 | 21.1 | 32.9 | 35.1 | 100.0 |
| A2/# | -23.1 | -21.4 | -19.5 | -15.4 | -20.9 | -27.0 | 1.1 | 13.9 | 14.5 | 18.5 | 100.0 | 35.1 |
| A3/# | -33.8 | -19.7 | -53.6 | -55.9 | -17.8 | -15.1 | 6.1 | 20.9 | 33.1 | 100.0 | 18.5 | 32.9 |
| T3/# | -25.0 | -12.2 | -50.9 | -56.0 | -16.7 | -18.7 | 6.4 | 24.9 | 100.0 | 33.1 | 14.5 | 21.1 |
| T1/# | -21.0 | -9.9 | -25.2 | -22.8 | -30.6 | -17.1 | 4.6 | 100.0 | 24.9 | 20.9 | 13.9 | 16.6 |
| T2/# | -10.3 | -13.0 | -6.6 | -6.3 | -1.3 | -7.2 | 100.0 | 4.6 | 6.4 | 6.1 | 1.1 | 5.5 |
| G2/# | 24.9 | 11.3 | 23.0 | 14.0 | 11.9 | 100.0 | -7.2 | -17.1 | -18.7 | -15.1 | -27.0 | -20.6 |
| G1/# | 12.4 | 12.4 | 18.6 | 17.5 | 100.0 | 11.9 | -1.3 | -30.6 | -16.7 | -17.8 | -20.9 | -24.3 |
| C3/# | 29.0 | 17.0 | 44.3 | 100.0 | 17.5 | 14.0 | -6.3 | -22.8 | -56.0 | -55.9 | -15.4 | -26.5 |
| G3/# | 32.9 | 15.3 | 100.0 | 44.3 | 18.6 | 23.0 | -6.6 | -25.2 | -50.9 | -53.6 | -19.5 | -29.2 |
| C2/# | 25.5 | 100.0 | 15.3 | 17.0 | 12.4 | 11.3 | -13.0 | -9.9 | -12.2 | -19.7 | -21.4 | -31.7 |
| C1/# | 100.0 | 25.5 | 32.9 | 29.0 | 12.4 | 24.9 | -10.3 | -21.0 | -25.0 | -33.8 | -23.1 | -45.1 |

C: Significance of correlation. - sign was ad log (-0) was regarded to be 100.

The sum of both complementary codon pairs (A+T and G+C) in every codon position is positively correlated to the sum of the same codon pair in the other two codon positions (Table 3). These correlations are valid for every species.

**Table 3**
**POSITIONAL NUCLEOTIDE USAGE FREQUENCIES IN 113 SPECIES**

| log(-C) | C1+G1 | C3+G3 | C2+G2 | C2+T2 | C1+T1 | G2+T2 | C3+T3 | G1+T1 | G3+T3 | A3+C3 | A1+C1 | A3+G3 | A2+C2 | A1+G1 | A2+G2 | A2+T2 | A3+T3 | A1+T1 |
|---|---|---|---|---|---|---|---|---|---|---|---|---|---|---|---|---|---|---|
| A1+T1 | -100 | -38.55 | -38.26 | -8.9 | -7.38 | -4.54 | -3.64 | -1.92 | -0.52 | 0.52 | 1.92 | 3.64 | 4.54 | 7.38 | 8.9 | 38.26 | 38.55 | 100 |
| A3+T3 | -38.55 | -100 | -24.89 | -4.64 | -6.94 | -2.71 | -4.02 | -0.44 | -1.96 | 1.96 | 0.44 | 4.02 | 2.71 | 6.94 | 4.64 | 24.89 | 100 | 38.55 |
| A2+T2 | -38.26 | -24.89 | -100 | -7.09 | -12.81 | -2.94 | -2.26 | -1.08 | -0.22 | 0.22 | 1.08 | 2.26 | 2.94 | 12.81 | 7.09 | 100 | 24.89 | 38.26 |
| A2+G2 | -8.9 | -4.64 | -7.09 | -100 | -3.01 | -7.09 | -5.41 | -11.21 | -0.45 | 0.45 | 11.21 | 5.41 | 7.09 | 3.01 | 100 | 7.09 | 4.64 | 8.9 |
| A1+G1 | -7.38 | -6.94 | -12.81 | -3.01 | -100 | -0.22 | -3.73 | -0.17 | -0.71 | 0.71 | 0.17 | 3.73 | 0.22 | 100 | 3.01 | 12.81 | 6.94 | 7.38 |
| A2+C2 | -4.54 | -2.71 | -2.94 | -7.09 | -0.22 | -100 | -0.78 | -2.36 | -0.17 | 0.17 | 2.36 | 0.78 | 100 | 0.22 | 7.09 | 2.94 | 2.71 | 4.54 |
| A3+G3 | -3.64 | -4.02 | -2.26 | -5.41 | -3.73 | -0.78 | -100 | -1.26 | -0.97 | 0.97 | 1.26 | 100 | 0.78 | 3.73 | 5.41 | 2.26 | 4.02 | 3.64 |
| A1+C1 | -1.92 | -0.44 | -1.08 | -11.21 | -0.17 | -2.36 | -1.26 | -100 | -0.6 | 0.6 | 100 | 1.26 | 2.36 | 0.17 | 11.21 | 1.08 | 0.44 | 1.92 |
| A3+C3 | -0.52 | -1.96 | -0.22 | -0.45 | -0.71 | -0.17 | -0.97 | -0.6 | -100 | 100 | 0.6 | 0.97 | 0.17 | 0.71 | 0.45 | 0.22 | 1.96 | 0.52 |
| G3+T3 | 0.52 | 1.96 | 0.22 | 0.45 | 0.71 | 0.17 | 0.97 | 0.6 | 100 | -100 | -0.6 | -0.97 | -0.17 | -0.71 | -0.45 | -0.22 | -1.96 | -0.52 |
| G1+T1 | 1.92 | 0.44 | 1.08 | 11.21 | 0.17 | 2.36 | 1.26 | 100 | 0.6 | -0.6 | -100 | -1.26 | -2.36 | -0.17 | -11.21 | -1.08 | -0.44 | -1.92 |
| C3+T3 | 3.64 | 4.02 | 2.26 | 5.41 | 3.73 | 0.78 | 100 | 1.26 | 0.97 | -0.97 | -1.26 | -100 | -0.78 | -3.73 | -5.41 | -2.26 | -4.02 | -3.64 |
| G2+T2 | 4.54 | 2.71 | 2.94 | 7.09 | 0.22 | 100 | 0.78 | 2.36 | 0.17 | -0.17 | -2.36 | -0.78 | -100 | -0.22 | -7.09 | -2.94 | -2.71 | -4.54 |
| C1+T1 | 7.38 | 6.94 | 12.81 | 3.01 | 100 | 0.22 | 3.73 | 0.17 | 0.71 | -0.71 | -0.17 | -3.73 | -0.22 | -100 | -3.01 | -12.81 | -6.94 | -7.38 |
| C2+T2 | 8.9 | 4.64 | 7.09 | 100 | 3.01 | 7.09 | 5.41 | 11.21 | 0.45 | -0.45 | -11.21 | -5.41 | -7.09 | -3.01 | -100 | -7.09 | -4.64 | -8.9 |
| C2+G2 | 38.26 | 24.89 | 100 | 7.09 | 12.81 | 2.94 | 2.26 | 1.08 | 0.22 | -0.22 | -1.08 | -2.26 | -2.94 | -12.81 | -7.09 | -100 | -24.89 | -38.26 |
| C3+G3 | 38.55 | 100 | 24.89 | 4.64 | 6.94 | 2.71 | 4.02 | 0.44 | 1.96 | -1.96 | -0.44 | -4.02 | -2.71 | -6.94 | -4.64 | -24.89 | -100 | -38.55 |
| C1+G1 | 100 | 38.55 | 38.26 | 8.9 | 7.38 | 4.54 | 3.64 | 1.92 | 0.52 | -0.52 | -1.92 | -3.64 | -4.54 | -7.38 | -8.9 | -38.26 | -38.55 | -100 |

C: Significance of correlation. - sign was added to negative correlations.

This strong positional correlation between codon bases suggests that it is possible to predict the frequency of usage of a nucleotide in the codon usage table from the frequencies of other nucleotides. Predictions regarding the third nucleotides in codons are especially interesting, because these are wobble bases for most amino acid codons.

I used the correlation between the sum of complementary codon pairs in the 1$^{st}$ and 2$^{nd}$ codon positions to predict the wobble bases using the frequencies for 113 different species (Table 4, Figure 7).

**Table 4**
**WOBBLE BASE PREDICTION**

| | | | | | | | | |
|---|---|---|---|---|---|---|---|---|
| A3/# | = | A1+T1/# | x | 1.1004 | + | -0.2898 | p= | 9.3E-38 |
| A3/# | = | A2+T2/# | x | 1.39228 | + | -0.6003 | p= | 8.6E-25 |
| C3/# | = | C1+G1/# | x | 1.14274 | + | -0.3466 | p= | 8.7E-34 |
| C3/# | = | C2+G2/# | x | 1.42309 | + | -0.3182 | p= | 8.1E-22 |
| G3/# | = | C1+G1/# | x | 0.95213 | + | -0.2566 | p= | 4.1E-38 |
| G3/# | = | C2+G2/# | x | 1.23154 | + | -0.251 | p= | 8.5E-27 |
| T3/# | = | C1+G1/# | x | 0.99447 | + | -0.2019 | p= | 7.2E-30 |
| T3/# | = | C2+G2/# | x | 1.26235 | + | -0.4851 | p= | 5.5E-21 |

This is of course a prediction of the frequencies of the four wobble bases in all 64 possible codons and has no predictive value for individual wobble bases belonging to individual amino acids.

All these correlation were of course carefully compared to corresponding random controls. Care was taken to ensure that the randomized control samples had the same size and distribution as the test samples. The sum of randomized fractions was kept equal to 1, as in the test samples. There were no correlations between the corresponding nucleotides in the control samples.

**Figure 7**

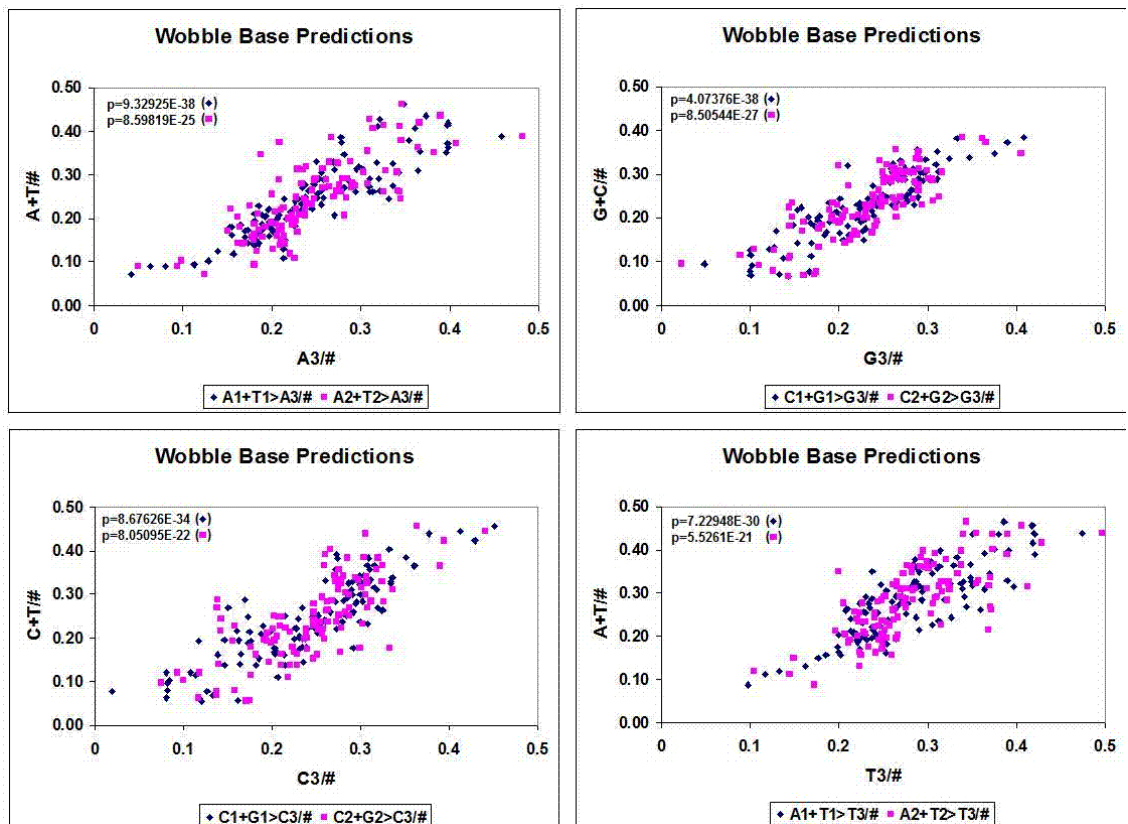

**Figure 7.: Correlation between Codon Bases in Codons of 113 Species**
The frequency of the four possible nucleotide bases (A, T, G, G) in the 3 possible codons positions ($1^{st}$, $2^{nd}$, $3^{rd}$) were counted in 113 codon usage tables and plotted against each other. A1+T1>A3 means the correlation between the sum of the $1^{st}$ A plus $1^{st}$ T frequencies and the $3^{rd}$ A frequency (n=113).

This simple but highly significant and species-independent positional relationship between NUFs provides further strong support for the view that the genetic code is the result of development and not at all a "frozen accident".

*Correlation between individual codons*

The detection of a strong internal pangenomic relationship among codons in the CUF Tables and the positional correlation among the base residues of these codons led to an even deeper correlation analysis. The correlations between every single codon frequency and every other codon frequency (64x64/2=2,048) were calculated using linear regression analysis [**Supplementary File 5**].

Further detailed analysis of the internal positional correlations between codons and codon bases revealed significant correlations between different codons, which are generally valid for every species in our collection.

I noticed that there is a pattern of positive/negative correlations in these tables corresponding to the codon letters and their positions in the codon. The general rules of this pattern are summarized in Figure 8.

...

**Figure 8**

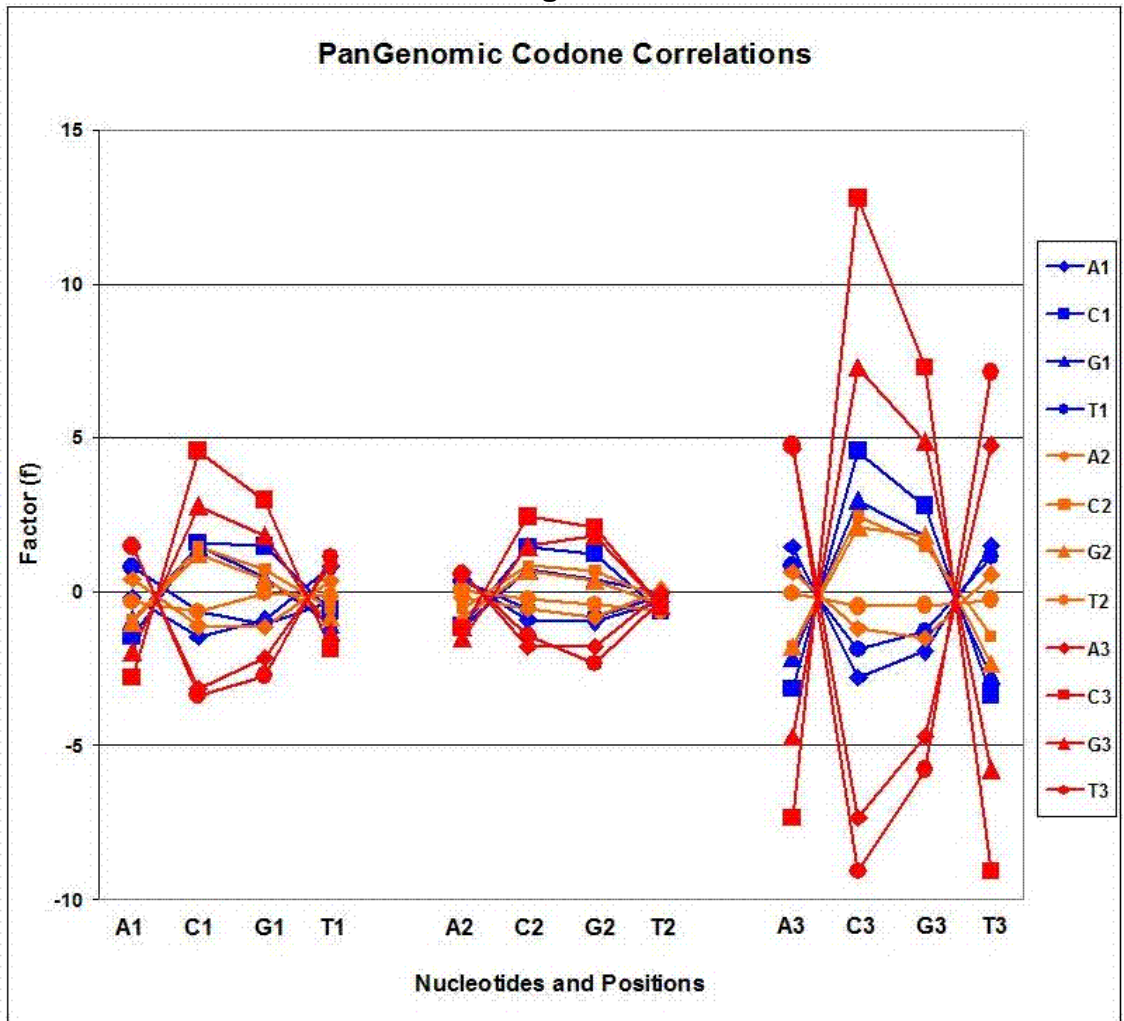

**Figure 8.: CUF - Pan-Genomic Codon Correlations.**
Codon frequencies were collected from 113 Codon Usage Frequency Tables and the correlation coefficients (C, 64x64) were calculated. f = -log C. A – sign was added to indicate negative correlations. The figure shows the f values between 4x4 codon letter combinations in 3x3 codon positions. Each symbol represent the mean of f values (n=113). f<-2 and f>2 correspond to statistically significant correlations.

There is a simple rule regarding codon correlations in the pangenome: there are positive correlations between complementary nucleotides and negative correlations between non-complementary nucleotides. This pattern of correlations is statistically significant in most combinations of nucleotide positions in codons. The correlations are statistically most significant between nucleotides in the 3$^{rd}$ codon positions.

*Prediction of individual wobble bases*

I used these correlations to predict individual wobble bases (all 64) from the 1st and 2nd letters of the codons (all 64). The possible correlations between a codon and the 16 possible permutations of the 4 1st and 2nd codon letters (64x4x4=1024) are listed in [**Supplementary File 6**].

*Accuracy of codon predictions*

I used the strongest correlations [Supplementary File 6] to predict codon frequencies, and the mean of several predictions was used as the averaged predicted value (p). Four different approaches were used to evaluate the predictions quantitatively.

The correlation between real (r) and predicted (p) values belonging to the same codons was significant ($p<0.05$) in 54 cases but not the other 10 (Figure 9a).

The correlation between real (r) and predicted (p) values belonging to the same species was significant ($p<0.05$) in all 113 cases and The p value was below 10E-07 in all but 2 species (Figure 9b).

**Figure 9**

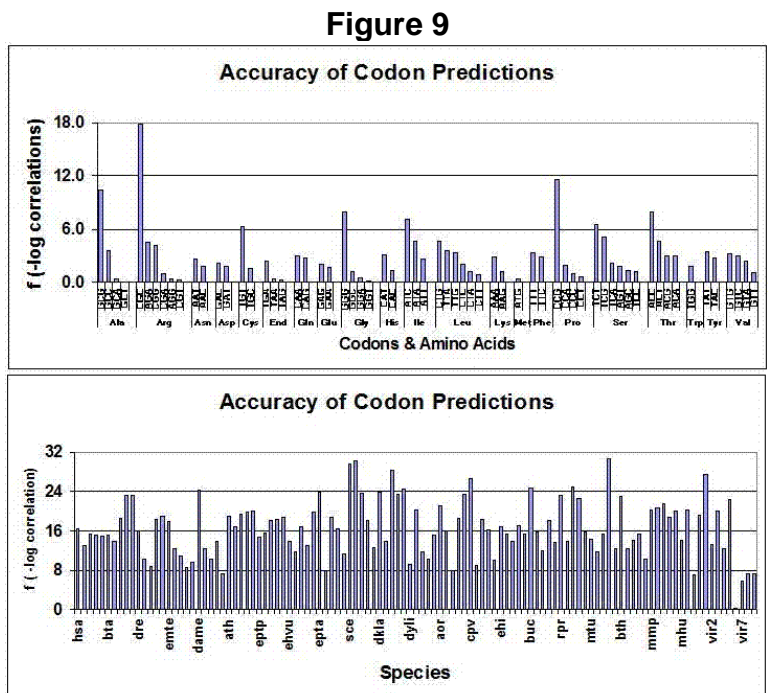

**Figure 9.: Accuracy of Codon Predictions - Amino Acid and Species Related Predictions**
Codon frequencies (64) were predicted (p) in 113 species and compared to the real (r) values. The correlations between r and p were sorted for codons (upper part) and species (lower pert). The correlations were expressed as f values (-log correlation coefficient). An f>1.5 can be regarded as statistically significant correlation.

The average accuracy of individual CUF predictions in 113 species and 87 individual proteins was estimated by comparing the average real and predicted frequencies. The significance of the correlation between real and predicted CUF was 1.3E-64 when data from 113 species were averaged and compared (n=64) and 1.9E-28 when data derived from 87 individual proteins (n=64) were used (Figure 10).

### Figure 10

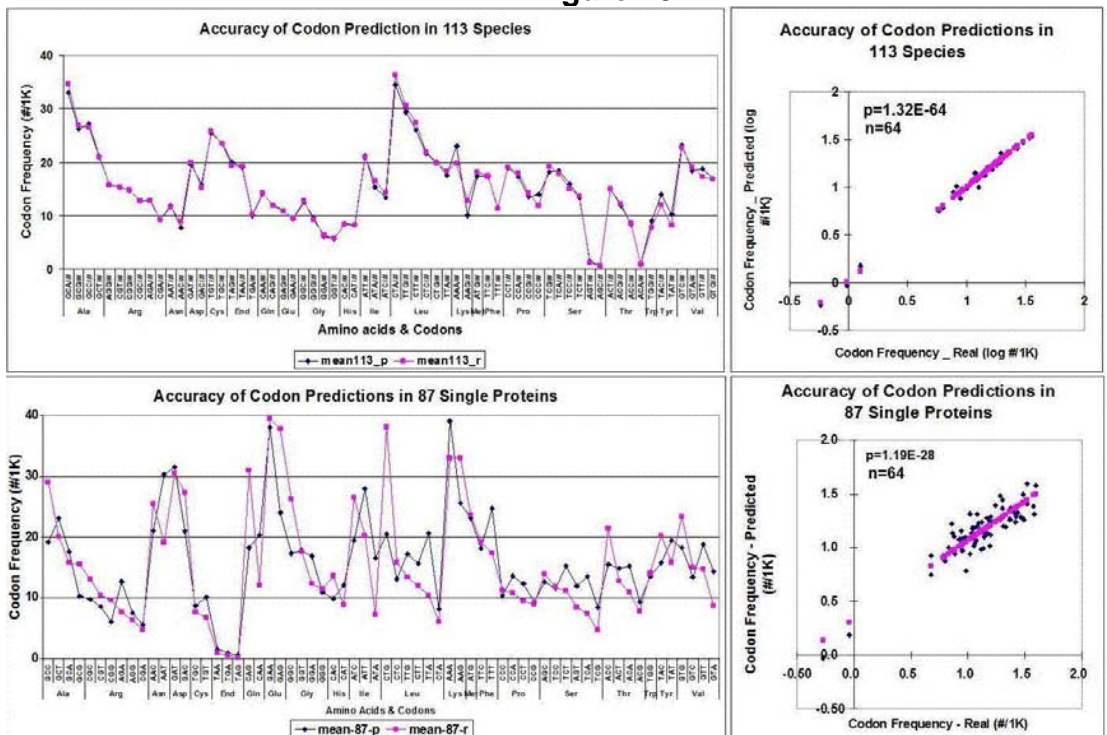

**Figure 10.: Accuracy of Codon Predictions in species and proteins**
Codon frequencies were predicted in 113 species and in 87 individual proteins. The average real (r) and predicted (p) codon frequencies were plotted (left) and correlations were analyzed (right).

## Discussion

There are basically two approaches to measuring CUB. First, relative synonymous codon usage (RSCU) values can be calculated [5]. RSCU is the observed number of codon occurrences divided by the number expected if synonymous codons were used uniformly. Second, the relative merits of different codons can be assessed from the viewpoint of translational efficiency. This second approach led to the development of the Codon Adaptation Index (CAI, [6]). The CAI model assigns a parameter, termed `relative adaptiveness', to each of the 61 codons (stop codons excluded). The relative adaptiveness of a codon is defined as its frequency relative to the most often-used synonymous codons and is computed from a set of highly expressed genes. The CAI is widely used even though the subjectivity involved in selecting the reference codons is well recognized [26, 27].

My way of calculating CUB is very close to the original suggestion [5] and regards uniform codon usage as the "null hypothesis"; any deviation from this is the bias. This approach made it possible to avoid subjectivity and species limitations in choosing the reference set of codons, and I can build the concept of CUB on the massive foundation of statistical laws and the large collection of sequence data collected in Codon Usage Frequency Tables.

The origin and biological significance of CUB is not well understood, therefore I tried to find the rules (if any) of its evolutionary development and gain new insights about its possible function. I sort my findings into two main categories: I found a.) some (few) signs of the evolutionary origin and development of CUB;

b.) unexpectedly large number of highly significant intern correlations between different codon residues (bases) at different codon positions (first, central, wobble) as well as between individual codons.

Inter-species variation in CUB is about 10%, but it is obvious that prokaryotes have significantly larger CUBs than eukaryotes. Bacteria may show the greatest bias because these primitive organisms are rich in highly-expressed genes and often use only one *dominant* codon. CUB decreases progressively with evolution and humans have the lowest bias (only about 20%). Evolutionary increase in codon number and genome complexity seems to reduce the CUB. It is noticeable that the average CUB (29.3 ± 1.1% (S.E.M.) n=113) means that synonymous codon usage frequencies are 29.3% distant from the *"all codons are equally good"* hypothesis, and 70.7% distant from the *"one codon is the best `codon"* alternative.

A more detailed qualitative analyzes of CUB is possible using a pan-genomic CUF Table. The original purpose of this virtual table was to create a reference for comparison of CUBs, but it turned out to reveal other codon-related connections too. The pan-genomic CUF Table is based on only 113 species, so it might be the first but not the last of its kind. It makes it possible to detect major, universal trends in codon usage behind small individual (or even species-wide) variations.

CUB is often correlated to the intensity of translation and has even been used to predict highly-expressed genes [6]. It is also known to be related to tRNA copy number, and co-evolution of tRNA gene composition and codon usage bias in genomes has been suggested [28]. I found a very strong correlation between the number of synonymous codons and the frequency of the amino acids they encoded, as well as the CUB. More synonymous codons encode more amino acids of the same kind and cause greater bias. This (rather logical) connection is not described in the literature, probably because the definition of CUB is very different from mine.

I tried to define a kind of "phylogenetic tree" of CUBs using the pan-genomic CUF table as reference. The significance of correlations between species-specific CUF and pan-genomic CUF gave a qualitative, theoretical measure of distances between codon usages. However this correlation-based approach did not successfully detect any recognizable, species-related evolutionary pattern.

Estimation of codon commitments through evolution showed that some codons are clearly over-utilized while other are avoided in most species. This finding is compatible with the concept of *dominant* and *suppressed* codons, but without stating that this difference is the result of evolution [29].

The non-randomness of synonymous codon usage is widely accepted today, and it has been suggested that independent forces (such as tRNA pool size [30]) have a role in the reading frame and there are contextual constraints on synonymous codon choice [31-33].

Other lines of evidence suggest that the Genetic Code itself (the 64 codons *in toto* as a system) has an inherited, internal structure [34, 35]. Statistical studies on the nucleotide compositions of codons and of different codon positions support this concept [36-41].

I searched for the origin end development of codon bias and I found an extensive network of internal correlations between codons of a species and the nucleotides that define them. The correlations described in this article are:

- Correlation between the frequency of any single codon residue (base) at any codon position (first, central, wobble) *and* the frequency of any other single codon residue (base) at any other codon position (also first, central, wobble);

- Correlation between the sum of frequencies of any two codon residues (bases) at any two codon positions ((first, central, wobble) *and* the sum of any two other codon residues (bases) at any two other codon positions (also first, central, wobble);

- Correlation between A+T, G+C frequencies at the $1^{st}$, $2^{nd}$ codon positions *and* A+T, G+C frequencies at the $3^{rd}$ codon position;

- Correlations between any two codons.

There seems to be a simple rule behind all these statistically significant correlations: *the correlation between any two nucleotides at any two codon positions is positive if the two nucleotides are complementary to each other and negative if they are not* (illustrated in Figure 8).

The large number of statistically highly significant correlations made it possible to predict the frequencies of synonymous codons (in 113 species and 87 individual proteins) from the general overall frequencies of codons. The reliability of predictions was tested.

**Conclusions**

The cumulative Codon Usage Frequency of any codon is strongly dependent on the cumulative Codon Usage Frequency of other codons belonging to the same species. The rules of this codon dependency are the same for all species and reflect WC base pair complementarity. This internal connectivity of codons indicates that all synonymous codons are integrated parts of the Genetic Code with equal importance in maintaining its functional integrity. The so-called codon bias is a bias caused by the protein-centric view of the genome.

**Acknowledgement**

The continuous support and editorial help of Dr P S Agutter is very much acknowledged.

...

**Additional files provided with this submission:**
**Legends to the Supplements**

Additional file 1: birosuppl 1_cuf tables_summary of 113 species.xls, 154K
http://www.tbiomed.com/imedia/1057595850210811/supp1.xls
**Suppl 1.: CUB Tables - Summary of 113 Species**

Additional file 2: birosuppl 2_calculation of codon usage bias (cub) - explanation , 43K
http://www.tbiomed.com/imedia/2110242038210811/supp2.xls
**Suppl 2. Calculation of Codon Usage Bias (CUB) – Explanation and Example**

The 64 codons were sorted in to 21 subgroups (fractions) corresponding to the 20 coded amino acids and the stop signal. The sum of synonymous codon frequencies were always regarded as 100% i.e. the sum of all codon frequencies is 2100% (color coded columns).

The fractional frequency ($CUF_{ij}$ %) of a synonymous codon is the contribution of that codon to this 100%. The theoretical, natural frequencies of the synonymous codons is regarded as equal to each other (for example the natural fractional frequency of each synonymous codon of Arg is 100%/6 =16.7%). The difference between this theoretical (calculated) frequency and the real (counted) fractional frequency of a codon is the $CUB_{ij}$ %. However it is necessary to use the |$CUB_{ij}$ %| value instead to be able to calculate and compare the total CUB values of entire proteins (i.e. the sum of 64 CUB values).

A theoretical extreme case of codon usage is when only one of all synonymous codons is used (CUB% 1 max column). The maximal possible CUB of all codons will in this case be 2416.7%, which is regarded as the CUFmax.

In the real case of *Homo sapiens* the sum of fractional frequencies is 456, which is 18.9% of the theoretical CUBmax.

Additional file 3: birosuppl 3_correlation analyses of codon usage bias (cub) in 11, 502K
http://www.tbiomed.com/imedia/1580949298210811/supp3.xls
**Suppl 3. Correlation Analyses of Codon Usage Bias (CUB) in 113 Species.**
CUBs of 113 species (each containing 64 values) were compared to the virtual CUB values in the Pan-Genomic Codon Usage Table by linear regression analyses. The - log C values were used as a measure of similarity and are indicated by horizontal bars at the right edge of the table. C: significance of correlation. The subgroups, corresponding to larger phylogenetic categories, are color coded and mean values for the groups are also indicated. The numbers of species in the subgroups are given in the "Mean" rows.

Additional file 4: birosuppl 4_cub committment and variation.xls, 119K
http://www.tbiomed.com/imedia/2423885362108116/supp4.xls
**Suppl Table 4 – CUB commitment and variation**
The 64 codons in 113 species were sorted according to the size and +/- orientation of their CUB. Some manual adjustments were made to segregate the data into four approximately symmetrical subgroups (corresponding to the color codes).

Additional file 5: birosuppl 5_ cuf_pangenomic codon correlations.xls, 882K
http://www.tbiomed.com/imedia/1300236762210811/supp5.xls
**Suppl 5. CUF- Pan-Genomic Codon Correlations.**
Codon frequencies were collected from 113 Codon Usage Frequency Tables and the significances of correlations (C, 64x64) were calculated. (n=113). The table displays the –log C values. A – sign was added to the -log C value to indicate negative correlations. Significant positive correlations (values >2) are indicated by **bold** numbers and gray background, while significant negative correlations (values < -2) are indicated by *italic* numbers and pink background. The collected data are sorted into 4x4x3x3=144 different subgroups corresponding to the 4x4 codon letter combinations and the 3x3 codon positions (red letters).

Additional file 6: birosuppl 6_cuf_wobble predictions.xls, 265K
http://www.tbiomed.com/imedia/4264769482108116/supp6.xls
**Suppl. 6 Prediction of Wobble Bases**
List of correlations between the frequency of a codon and the frequency of other codons, which contain the 4x4 permutations of codons at the 1$^{st}$ and 2$^{nd}$ codon positions. 64 times 16 equations were calculated from these correlations. Only the strongest correlations, those used in codon predictions, are listed and color coded. Positive correlations are indicated by bold letter in blue background and negative correlations are given by italic letters in pink background. F is as defined in fig. 9.